\documentstyle[preprint,aps,epsfig]{revtex}
\preprint{\vbox{  
\hbox{IFT-P.013/03}   
\hbox{April 2003} 
}}  
\begin{document}  
\draft 
\title{Mapping the Wigner distribution function of the Morse oscillator into a
semi-classical distribution function
}
\author{ G. W. Bund$^1$~\footnote{electronic address: bund@ift.unesp.br} 
and M. C. Tijero$^{1,2}$~\footnote{electronic address: maria@ift.unesp.br} } 
\address{
$^1$~Instituto de F\'\i sica  Te\'orica\\ 
Universidade  Estadual Paulista\\
Rua Pamplona, 145\\ 
01405-900-- S\~ao Paulo, SP\\ 
Brazil
\\$^2$ 
Pontif\'\i cia Universidade Cat\'olica de S\~ao Paulo \\
Rua Marqu\^es de Paranagu\'a, 111 \\01303-000-- S\~ao Paulo, SP\\ 
Brazil} 
\maketitle 
\newpage
\begin{abstract}  
The mapping of the Wigner distribution function (WDF) for a given bound-state
onto a semiclassical distribution function (SDF) satisfying the Liouville
equation introduced previously by us is applied to the ground state of the Morse
oscillator. Here we give results showing that the SDF gets closer to the
corresponding WDF as the number of levels of the Morse oscillator increases. We
find that for a Morse oscillator with one level only, the agreement between the
WDF and the mapped SDF is very poor but for a Morse oscillator of ten levels it
becomes satisfactory. 
\end{abstract}
\pacs{PACS   numbers: 03.65.-w; 
  03.65.Sq 
}

\section{Introduction}
\label{sec:intro}

The Wigner description in phase-space~\cite{wigner} provides a tool for
comparing quantum and classical dynamics~\cite{lees}.
In a previous paper~\cite{gwb1} we introduced a mapping relating the Wigner
distribution function (WDF) corresponding to a given wave function, solution of
the Schr\"odinger equation to a semiclassical distribution function (SDF)
satisfying the classical Liouville equation with the same potential. So far this
mapping was applied to the ground state of the square well~\cite{gwb1}, the
P\"oschl-Teller potential~\cite{gwb2} and to a modified harmonic oscillator
potential~\cite{gwb2}.

Here this mapping is extended to the ground state of the unidimensional Morse
oscillator. We vary the depth of the potential in order to study the effect of
the level density on the mapping.

The Morse oscillator~\cite{morse} plays an important role in many areas of
Physics, because it is a good approximation to diatomic molecular
potentials~\cite{jpd1}. Today the Morse potential is used in molecular
spectroscopy, even for polyatomic molecules~\cite{nova}. Some authors have used the 
Morse potential for studying the semiclassical limit of Quantum
Mechanics~\cite{jpd2}.

The unidimensional Morse oscillator in phase-space has been already studied by many authors.
The classical motion~\cite{marcus} as well as the quantum
representation~\cite{jpd1} are well known for this potential.

In Sec.~II of this work we summarize the method developed in Ref.~\cite{gwb1}
for mapping WDF into SDF. In Sec.~III we study the Morse
oscillator and in Sec.~IV we present our results and conclusions.

\section{Description of the method}
\label{sec:ii}

In this section we present a short derivation of the method used for the
mapping~\cite{gwb1}. We start with the quantum Liouville equation for the WDF
$\rho$: 
\begin{equation}
\frac{\partial\rho}{\partial t}+\frac{p}{m}\frac{\partial\rho}{\partial q}+
\int K(q,p-p')\rho(q,p',t)dp'=0,
\label{e1}
\end{equation}
where the kernel $K$ is given by

\begin{equation}
K(q,p-p')=\frac{i}{\hbar}\int\frac{dv}{2\pi\hbar}
\exp\left[\frac{i}{\hbar}(p-p')v\right]
\left[V\left(q-\frac{v}{2}\right)-V\left(q+\frac{v}{2}\right)\right],
\label{e2}
\end{equation}
$V(q)$ being the potential. The corresponding classical Liouville equation may
be written similarly

\begin{equation}
\frac{\partial\rho_c}{\partial t}+\frac{p}{m}\frac{\partial\rho_c}{\partial q}+ 
\int K_c(q,p-p')\rho_c(q,p',t)dp'=0,
\label{e3}
\end{equation}
where
\begin{equation}
K_c(q,p-p')=-\frac{i}{\hbar}\int\frac{dv}{2\pi\hbar}
\exp\left[\frac{i}{\hbar}(p-p')v\right]\,v\,\frac{\partial V}{\partial q}
=-\frac{\partial V}{\partial q}\frac{\partial }{\partial p}\delta(p-p').
\label{e4}
\end{equation}

We relate solutions $\rho$ and $\rho_c$ of Eqs.~(\ref{e1}) and (\ref{e3})
through the integral equation
\begin{eqnarray}
\rho(q,p,t)&=&\rho_c(q,p,t)-\int^\infty_{-\infty} dt'\int dq'dp'
G_c(q,p;q',p',t-t') 
\nonumber \\ &\times&\int
dp^{\prime\prime}[K(q',p'-p^{\prime\prime})-K_c(q',p'-p^{\prime\prime})]
\rho(q',p^{\prime\prime},t'),
\label{e5}
\end{eqnarray}
where $G_c$ is the retarded Green's function corresponding to the classical
Liouville equation (\ref{e3}). In fact Eq.~(\ref{e5}) is the defining equation
of $\rho_c$ since the WDF $\rho$ is already determined by fixing the wave
function solution of the Schr\"odinger equation. Using the differential equation
satisfied by $G_c$ \cite{gwb1} one may easily verify that
$\rho_c$ satisfies Eq.~(\ref{e3}) provided $\rho$ obeys Eq.~(\ref{e1}).

As $K_c$ is the term of lowest order in the expansion of $K$ in a power series
of $\hbar$, $\rho-\rho_c$ contains only corrections of the first and higher order
terms in $\hbar$.
 
It can be shown from results given in Ref.~\cite{gwb1}, for a WDF generated from
solutions of the Schr\"odinger equation for which the initial wave function at
the time $t=t_0$ is given; assuming $\rho(q,p,t)$ to vanish for $t<t_0$, that
$\rho_c$  from Eq.~(\ref{e5}) gives simply the classical evolution of
$\rho(q,p,t_0)$, that is
\begin{equation}
\rho_c(q,p,t)=\rho(Q(q,p,t_0-t ),P(q,p,t_0-t ) ,t_0),\quad t\geq t_0.
\label{e6}
\end{equation}
Here $(Q(q,p,\tau),P(q,p,\tau))$ describes the classical
trajectory in phase-space of a particle subject to the potential $V(q)$ which at
the time $\tau=0$ occupies the point $(q,p)$, that is
\begin{equation}
q=Q(q,p,0),\quad p=P(q,p,0).
\label{e7}
\end{equation}

In what follows we shall consider mostly the mapping of the time independent
WDF $\rho(q,p)$ corresponding to a bound state. In this case, from Eq.~(\ref{e5})
one obtains~\cite{gwb1} the following prescription for $\rho_c$, 
\begin{equation}
\rho_c(q,p)=\lim_{\epsilon\to0_+}\epsilon\int^o_{-\infty}\,d\tau\,
e^{\epsilon\tau}\rho(Q(q,p,\tau ),P(q,p,\tau)),
\label{e8}
\end{equation}
which is also time independent. Here $e^{\epsilon\tau}$ is a convergence factor
introduced explicitly in the Green's function.

It can be shown~\cite{gwb1} that for points $(q,p)$ on closed classical orbits
the expression given by Eq.~(\ref{e8}) is equivalent to
\begin{equation}
\rho_c(q,p)=\frac{1}{T(q,p)}\int^0_{-T(q,p)}dt\,\rho(Q(q,p,t),P(q,p,t)), 
\label{e9}
\end{equation}
where $T=T(q,p)$ is the period of the orbit. For open trajectories we have
$T\to\infty$ and as for bound states the integral in Eq.~(\ref{e9}) converges, we
get $\rho_c=0$. As it will be discussed further, in the calculation of averages
of physical quantities the quantity of interest is rather $T(q,p)\rho_c(q,p)$
which does not vanish for the open trajectories.

We observe here that one may verify directly that the time-independent Liouville
equation is obeyed by applying the operator $p\partial/\partial q-(\partial
V/\partial q)\partial/\partial p$ on the right hand side of Eq.~(\ref{e8}) or
(\ref{e9}) and using the fact that $\rho(Q(q,p,-t),P(q,p,-t))$ satisfies the
classical Liouville equation (\ref{e3}).

An important property of the stationary SDF is that it is constant along the
classical trajectories~\cite{gwb1} which can be seen by noticing that for any
two points $(q,p)$ and $(q',p')$ on the same path in phase space one has a time
interval $\Delta$ such that
\begin{equation}
Q(q',p',t) = Q(q,p,t+\Delta),\quad P(q',p',t)=P(q,p,t+\Delta).
\label{e10}
\end{equation}
($\Delta$ is the time which the particle takes to go from the point  $(q,p)$ to
the point  $(q',p')$ in phase space or vice versa). Thus we get
\begin{eqnarray}
\rho_c(q',p')&=&\frac{1}{T}\int^0_{-T}dt\,\rho(Q(q',p',t),P(q',p',t))
\nonumber \\&=&
\frac{1}{T}\int^0_{-T}dt\,\rho(Q(q,p,t+\Delta),P(q,p,t+\Delta))
\nonumber \\ &=&\rho_c(q,p),
\label{e11}
\end{eqnarray}
where we used the periodicity of the integrand in the last step.

Let $(Q(E,\nu,t),P(E,\nu,t))$ represent, for $t$ in the interval 
$-T(E,\nu)/2\leq t\leq T(E,\nu)/2$ ($T$ may be infinity), the
points of a trajectory in phase-space corresponding to energy $E$ and fixed
value of $\nu$. Here the discrete parameter $\nu$ ($\nu=1,\cdots, \nu_{\rm
max}$) distinguishes  between the different trajectories corresponding to the
same value of $E$. By varying $E$ and $\nu$ one covers the entire allowed phase
space. Thus we may consider the transformation of points ($E,\nu,t$) in the
appropriate domains $D_\nu$ of the energy-time space onto the points ($p,q$) of
the phase space
\begin{equation}
q=Q(E,\nu,t),\quad p= P(E,\nu,t).
\label{e12}
\end{equation}

Let us define the function
\begin{equation}
R_c(E,\nu)=\int^{T(E,\nu)/2}_{-T(E,\nu)/2}dt\,\rho(Q(E,\nu,t),P(E,\nu,t)).
\label{e13}
\end{equation}

According to Eq.~(\ref{e9}) one has
\begin{equation}
\rho_c(q,p)=\frac{R_c(E(q,p),\nu)}{T(E(q,p),\nu)},
\label{e14}
\end{equation}
by choosing the value of $\nu$ appropriate to the trajectory containing the
point ($q,p$) as follows from Eq.~(\ref{e12}). We observe here that the
relationship (\ref{e13}) between $\rho(q,p)$ and $R_c(E,\nu)$ is analogous to
that between $R(q)$, the probability density in coordinate space and the WDF
$\rho(q,p)$~\cite{groot}:
\begin{equation}
R(q)=\int\rho(q,p)dp.
\label{e15}
\end{equation} 

The average of any Weyl function ~\cite{groot} ${\cal O}(q,p)$ corresponding to
a certain operator $O$ may be written
\begin{equation}
\langle O\rangle=
\int {\cal O}(q,p)\rho(q,p)dpdq=\sum_\nu\int_{\cal D_\nu}
o(E,\nu,\tau)r(E,\nu,\tau)dEd\tau,
\label{e16}
\end{equation}
where we introduced the functions
\begin{eqnarray}
o(E,\nu,\tau)={\cal O}(Q(E,\nu,\tau),P(E,\nu,\tau)), \\
r(E,\nu,\tau)=\rho(Q(E,\nu,\tau),P(E,\nu,\tau)),
\label{e17}
\end{eqnarray}
and we used the fact that the Jacobian $J$ of the transformation (\ref{e12}) is
unity~\cite{gwb1}
\begin{equation}
J=\left\vert\begin{array}{cc}
\frac{\partial P}{\partial E} & \frac{\partial Q}{\partial E} \\
\frac{\partial P}{\partial t} & \frac{\partial Q}{\partial t}
\end{array}\right\vert= \frac{\partial P}{\partial E} \frac{\partial
E}{\partial 
p}+  \frac{\partial Q}{\partial E} \frac{\partial E}{\partial q}=1.
\label{e18}
\end{equation}

In the special case in which the Weyl function ${\cal O}$ is a constant of
motion depending on $(q,p)$ only through the energy $E(q,p)$, we obtain from 
(\ref{e16}), (18) and (\ref{e13})
\begin{equation}
\langle O\rangle=
\sum_\nu\int_{D_\nu}
o(E,\nu)r(E,\nu,\tau)dEd\tau=\sum_\nu\int o(E,\nu)R_c(E,\nu)dE.
\label{e19}
\end{equation}

In particular the normalization condition for $R_c(E,\nu)$ 
\begin{equation}
\sum_\nu\int R_c(E,\nu)dE=1,
\label{e20}
\end{equation}
follows from Eq.~(\ref{e19}) by taking ${\cal O}(q,p)=1$ since the Wigner
function $\rho(q,p)$ is normalized.

As $\rho_c$ is an approximation correct in zeroth order of the expansion in
powers of $\hbar$ of the Wigner function $\rho$, the average
\begin{equation}
\langle O\rangle_c=\int{\cal O}(q,p)\rho_c(q,p)dqdp,
\label{e21}
\end{equation}
is also correct in the same order in $\hbar$. The expression (\ref{e21}) may be
also written, by making use of the transformation (\ref{e12}), as
\begin{eqnarray}
\langle O\rangle_c=\sum_\nu\int_{D_\nu}o(E,\nu,t)\frac{R_c(E,\nu)}{T(E,\nu)}dEdt=
\sum_\nu\int \bar{o}(E,\nu)R_c(E,\nu)dE,
\label{e22}
\end{eqnarray}
where we used Eqs.~(\ref{e14}) and (17) and $\bar{o}(E,\nu)$ denotes the
average
\begin{equation}
\bar{o}(E,\nu)=\int^{T(E,\nu)/2}_{-T(E,\nu)/2}o(E,\nu,t)\frac{dt}{T(E,\nu)}.
\label{e23}
\end{equation}

Thus the average $\langle O\rangle_c$ is obtained by replacing $o(E,\nu,t)$ in
Eq.~(\ref{e16}) by the average (\ref{e23}). If, for fixed $E$, $o(E,\nu,t)$
depends weakly on $t$, the average  $\langle O\rangle_c$ is expected to be a
good approximation.

\section{ Wigner distribution function for the Morse potential}
\label{sec:wdfmp}

In 1929 Morse~\cite{morse} suggested the potential $U_0(1-e^{-\alpha r})^2$
for studying diatomic molecules. The Schr\"odinger equation for this potential
does not have an exact solution, but for the one dimensional case the problem
can be solved analytically~\cite{nieto,haar}. 

In order to obtain the Wigner Distribution Function (WDF)~\cite{wigner}
\begin{equation}
\rho(q,p,t)=(2\pi\hbar)^{-1}\int^\infty_{-\infty}dv\,e^{\frac{i}{\hbar}pv}
\psi(q-\frac{1}{2}v,t)\psi^*(q+\frac{1}{2}v,t),
\label{wdf}
\end{equation}
where $\psi$ is the solution of the Schr\"odinger equation, we need the
eigenfunctions for the one-dimensional Morse potential 
\begin{equation}
{\cal V}(x)=D(1-e^{-a x})^2,
\label{mp}
\end{equation}
$a$ and $D$ being constant parameters. Starting with the Schr\"odinger equation 
\begin{equation}
\left[-\frac{\hbar^2}{2m}\frac{d^2}{dx^2}+D(1-e^{-a
x})^2\right]\psi_n={\cal E}_n\psi_n, 
\label{se1}
\end{equation}
we introduce the dimensionless parameter $\lambda$
\begin{equation}
\lambda=\frac{\sqrt{2mD}}{a\hbar},
\label{lambda}
\end{equation}
and the dimensionless coordinate
\begin{equation}
q=ax,
\label{ql}
\end{equation}
obtaining an eigenvalue equation depending only on one parameter
\begin{equation}
\left[-\frac{1}{\lambda^2}\frac{d^2}{dq^2}+(1-e^{-q})^2 \right]\psi_n=
\epsilon_n\psi_n,
\label{se2}
\end{equation}
where 
\begin{equation}
\epsilon_n=\frac{{\cal E}_n}{D}.
\label{ep}
\end{equation}

This equation is solved most conveniently using the variable
\begin{equation}
\xi=2\lambda \,e^{-q},\quad -\infty<q<\infty.
\label{xi}
\end{equation}

The eigenfunctions and eigenvalues are~\cite{jpd1}
\begin{eqnarray}
\psi_n(\xi)=N(n,\lambda)\,e^{-\frac{\xi}{2}}\,
\xi^{\lambda-n-\frac{1}{2}}L^{\lambda-n-\frac{1}{2}}_n(\xi),\\
\epsilon_n=\frac{2}{\lambda}\,\left(n+\frac{1}{2}\right)\,
\left[1-\frac{1}{2\lambda}\left(n+\frac{1}{2}\right)
\right], 
\label{pen}
\end{eqnarray}
where the quantum number $n$ takes the values
\begin{equation}
n=0,1,2,...[\lambda-\frac{1}{2}].
\label{n}
\end{equation}
Here $[\lambda-1/2]$ denotes the largest integer smaller than $\lambda-1/2$,
$L^s_n$ is the polynomial~\cite{jpd1}
\begin{equation}
L^s_n=\sum^n_{j=0}\left( 
\begin{array}{c}
n+2s\\
n-j\end{array}
\right)\,\frac{(-\xi)^j}{j!},
\label{ln}
\end{equation}
and the normalization factor $N$ is given by
\begin{equation}
N(\lambda,n)=\left[
\frac{(2\lambda-2n-1)\Gamma(n+1)}{\Gamma(2\lambda-n)}\right]^{\frac{1}{2}},
\label{nf}
\end{equation}
where the normalization condition $\int\psi^*_n\psi_ndq=1$ is assumed. In
Appendix A, Eqs.~(33) and (\ref{pen}) are derived. 

From Eq.~(\ref{pen}) one verifies that, for $\lambda\gg1$ and $n\ll\lambda$, the
energy spectrum of the Morse oscillator is written approximately ${\cal
E}_n\approx \hbar\omega_0(n+1/2)$, which is the spectrum of a harmonic
oscillator with frequency
\begin{equation}
\omega_0=\frac{2D}{\hbar\lambda}=a\sqrt{\frac{2D}{m}}=\lambda\,\hbar\,\frac{a^2}{m}.
\label{w0}
\end{equation}

As the semiclassical distribution functions $\rho_c$ for the harmonic oscillator
coincide with the Wigner functions $\rho$ we expect that if $\lambda\gg1$
$\rho_c$ does get close to $\rho$ for the low lying levels. Thus it may be
appropriate to use, instead of the variable $q$ and its canonical conjugate
momentum $p$, the variables which treat harmonic oscillators on the same
footing, namely~\cite{jpd1}
\begin{mathletters}
\label{39}
\begin{equation} 
Q=\left[\frac{m\omega_0}{\hbar}\right]^{\frac{1}{2}}x
=\left[\frac{2mD}{a^2\hbar^2\lambda}\right]^{\frac{1}{2}}q
=\sqrt{\lambda}q,
\label{39a}
\end{equation}
\begin{equation}
P=\frac{1}{\sqrt\lambda}p.
\label{39b}
\end{equation}

The coordinate $Q$ and the momentum $P$ have been used in the figures of
Sec.~\ref{sec:con}. We define the dimensionless potential $V(Q)$
\begin{equation}
V(Q)=(\hbar\omega_0)^{-1}{\cal V}(x)=
\frac{\lambda}{2}\left(1-e^{-\frac{Q}{\sqrt\lambda}}\right)^2,
\label{39c}
\end{equation}
which is also used in the figures. For the energy levels there
we use similarly
\begin{equation}
E_n=(\hbar\omega_0)^{-1}{\cal E}_n=
\left(n+\frac{1}{2}\right)-\frac{1}{2\lambda}\left(n+\frac{1}{2}\right)^2.
\label{39d}
\end{equation}
\end{mathletters}

In the special case of our interest, namely the ground state, $n=0$, the wave
function is given by
\begin{equation}
\psi_0(\xi)=\left[\frac{2\lambda-1}{\Gamma(2\lambda)}
\right]^{\frac{1}{2}}\xi^{\lambda-\frac{1}{2}}\,e^{-\frac{\xi}{2}}.
\label{wf0}
\end{equation}
In order to calculate the Wigner function replace $\psi$ by
$\psi_n(2\lambda e^{-q})$ into Eq.~(\ref{wdf}), obtaining 
\begin{equation}
\rho^{(\lambda)}_n(q,p)=(2\pi\hbar)^{-1}\int^\infty_{-\infty}dv\,
\psi_n(2\lambda e^{[-q+\frac{v}{2}]})\psi^*_n(2\lambda e^{[-q-\frac{v}{2}]})
\,e^{\frac{i}{\hbar}pv}.
\label{wdf2}
\end{equation}
Introducing the new integration variable
\begin{equation}
\tau=e^{\frac{v}{2}},
\label{t}
\end{equation}
Eq.~(\ref{wdf2}) becomes
\begin{equation}
\rho^{(\lambda)}_n(q,p)=(\pi\hbar)^{-1}\int^\infty_0
\psi^*_n\left(\frac{\xi}{\tau}\right)\psi_n(\xi\tau)\tau^{2ip\hbar^{-1}}\,
\frac{d\tau}{\tau},
\label{wdf3}
\end{equation}
where $\xi$ is given by Eq.~(\ref{xi}).
Substituting $\psi_n$ in Eq.~(\ref{wdf3}) by the expression (33)
$\rho^{(\lambda)}_n$ becomes~\cite{jpd1}
\begin{eqnarray}
\rho^{(\lambda)}_n(q,p)=\left[ \frac{\pi\hbar}{2}
\right]^{-1}[N(\lambda,n)]^2\xi^{2\lambda-2n-1}
\sum^n_{r=0}\sum^n_{s=0}b(\lambda,n,r)b(\lambda,n,s)\xi^{r+s}
K_{s-r+2ip\hbar^{-1}}(\xi),
\label{yuca}
\end{eqnarray}
where
\begin{equation}
b(\lambda,n,j)=\left(\begin{array}{c}
2\lambda-n-1\\
n-j\end{array}\right)\frac{(-1)^j}{j!}
\label{yuca2}
\end{equation}
and $K_\nu(\xi)$ is defined by~\cite{jpd1} 
\begin{equation}
K_\nu(\xi)=\frac{1}{2}\int^\infty_0\,e^{ -\frac{\xi}{2}\,\left(\tau+
\frac{1}{\tau}\right)}\;
\tau^\nu\;\frac{d\tau}{\tau},
\label{k}
\end{equation}
$\nu$ being a complex variable. In the particular case in which $n=0$ we get
\begin{equation}
\rho^{(\lambda)}_0(q,p)=\left(\frac{\pi\hbar}{2} \right)^{-1}\,
\frac{2\lambda-1}{\Gamma(2\lambda)}\,\xi^{2\lambda-1}\,K_{2ip\hbar^{-1}}(\xi).
\label{r2}
\end{equation}

The numerical method we used to calculate the function $K_\nu(\xi)$ will be
given in Appendix B.

The symmetries obeyed by the function $K_\nu$ are
\begin{eqnarray}
K_{\nu^*}(\xi)=\frac{1}{2}\int^\infty_0e^{-\frac{\xi}{2}(\tau+\frac{1}{\tau})}
(\tau^\nu)^*\frac{d\tau}{\tau}=(K_\nu(\xi))^*,\\
K_{-\nu}(\xi)=\frac{1}{2}\int^\infty_0e^{-\frac{\xi}{2}(\tau+\frac{1}{\tau})}
\tau^{-\nu} \frac{d\tau}{\tau}=K_\nu(\xi),
\label{mais}
\end{eqnarray}
where the last step is obtained by making $\tau\to\tau^{-1}$. 
From Eqs.(49) and (48) we get
\begin{equation}
K_{-a+ib}=K^*_{a+ib}.
\label{boa}
\end{equation}
Thus Eq.~(\ref{yuca}) may be written
\begin{eqnarray}
\rho^{(\lambda)}_n(q,p)=\left[ \frac{\pi\hbar}{2}
\right]^{-1}[N(\lambda,n)]^2\xi^{2\lambda-2n-1}
\sum^n_{r=0}\sum^n_{s=0}b(\lambda,n,r)b(\lambda,n,s)\xi^{r+s}
Re\left(K_{s-r+2ip\hbar^{-1}}(\xi)\right),
\label{yuca3}
\end{eqnarray}
showing explicitly that $\rho^{(\lambda)}_n$ is real.

In order to calculate the semiclassical distribution function $\rho_c(q,p)$
given by Eq.~(\ref{e9})  or (\ref{e14}), we need the solution of the classical
equation of motion
\begin{equation}
m\ddot{x}=-D\frac{d}{dx}(1-e^{-ax})^2,
\label{em}
\end{equation}  
which has already been obtained exactly~\cite{marcus}.
Introducing the coordinate $q=ax$ and the variable $\theta$ given by
\begin{equation}
\theta=\omega_0t,
\label{te}
\end{equation} 
where $\omega_0$ is given by Eq.~(\ref{w0}), Eq.~(\ref{em}) becomes
\begin{equation}
2\frac{d^2q}{d\theta^2}=-\frac{d}{dq}(1-e^{-q})^2.
\label{em2}
\end{equation}
The solution of Eq.~(\ref{em2}) for $\epsilon=E/D<1$, where $E$ is the
energy associated with the trajectory, is
\begin{equation}
q(E,t)=\ln\left\{\frac{1}{1-\epsilon}\left[1+\sqrt{\epsilon} 
\sin\sqrt{1-\epsilon}(\omega_0t-\theta_0)\right]\right\}.
\label{sem2}
\end{equation} 

The canonical momentum associated with $q$ is
\begin{equation}
p=\frac{m\dot{q}}{a^2},
\label{mo}
\end{equation}
which using Eqs.~(\ref{sem2}) and (\ref{w0}) gives
\begin{equation}
p(E,t)=\hbar\lambda\,\frac{\sqrt{\epsilon}
\sqrt{1-\epsilon}\cos[\sqrt{1-\epsilon}(\omega_0t-\theta_0)]}
{1+\sqrt{\epsilon}\sin[\sqrt{1-\epsilon}(\omega_0t-\theta_0)]}.
\label{mo2}
\end{equation}

The period associated with the orbit is
\begin{equation}
T(E)=\frac{2\pi}{\omega_0\sqrt{1-\epsilon}}\,.
\label{pe}
\end{equation}

Analising Eq.~(\ref{sem2}) one obtains that for $E\ll D$ the trajectories are
close to those of a harmonic oscillator of frequency $\omega_0$.

\section{Results}
\label{sec:con}

In this section we present the results of our calculations. We calculated the
WDF $\rho$ and corresponding SDF $\rho_c$ for the ground state of the Morse
oscillator choosing for the parameter $\lambda$ the values 1,2,4 and 10
corresponding to oscillators with 1,2,4 and 10 levels respectively. 
In the figures we used throughout the dimensionless coordinates $Q$ and the
conjugate momenta $P$ (in units of $\hbar$) defined by Eqs.~(\ref{39a}) and
(\ref{39b}). In Figs. 1,2,3 and 4 the potentials $V(Q)$ defined by
Eq.~(\ref{39c}) are displayed for $\lambda=1,2,4$ and 10 and the energy levels
marked. We observe here that the potential given by Eq.~(\ref{39c}) is
independent of $\lambda$ for $Q^2\ll\lambda$.
 
Figs.~5,6,7 and 8 reproduce our calculations of the WDF for
$\lambda=1,2,4$ and 10 respectively through curves of constant density
$\rho(Q,P)$. The value of $\rho$ varies from $\rho\approx0.3$ to $\rho=0$ in the
region of phase-space corresponding approximately to the region of the bound
classical particles. Then there occur adjacent regions on
which $\rho$ alternates from negative to positive values and its 
magnitude decreases as the region gets farther away from the origin.

In the case $\lambda=1$ the minimum value of $\rho$ is (in units of
$\hbar^{-1}$) of the order of $-10^{-2}$, as it can be seen from Fig.~5. This minimum approaches
zero as $\lambda$ increases, which can be observed from Fig.~7 for $\lambda=4$
where the minimum of $\rho$ is about $-10^{-4}$. 

The maximum of $\rho$, as it can be verified from Figs.~5---8, moves from
the point $(Q,P)=(1.2,0)$ to $(Q,P)=(0.3,0)$ as 
$\lambda$ increases from 1 to 10 and its value increases slightly as $\lambda$
increases. Thus the curve for $\rho=0.3$, present for $\lambda\geq2$, does not
occur for $\lambda=1$. 

Another feature of the WDF is that the curves of constant density 
$\rho$ become more
symmetric with respect to an axis parallel to the $P-$axis as $\lambda$
increases, becoming close to the form of an ellipse. 
This tallies with the fact
that for $Q\ll\sqrt\lambda$, $V(Q)$ is the potential of a harmonic oscillator. 

In Figs.~10,11,12 and 13 we present SDF curves for fixed $\rho_c$ superposed on
WDF curves with $\rho=\rho_c$ for comparison.
It will be seen that, as $\lambda$ increases the SDF approximation improves,
which means also that the WDF curves of constant $\rho$ become closer to
classical trajectories. In Fig.~9 we plot curves of constant $\rho_c$ for the
case $\lambda=1$. For this oscillator our semiclassical approximation
is anomalous, as the value of $\rho_c$ increases from $\rho_c=0.145$ to
$\rho_c=0.179$ as the classical energy $E$ varies from $E=0$ to
$E=0.26\hbar\omega_0$ but decreases as $E$ increases
further. For the other oscillators $\rho_c$ decreases as the energy $E$
increases until reaching the value $\rho_c=0$. As a consequence for $\lambda=1$
one has two closed curves 
with the same $\rho_c$ for $0.145<\rho_c<0.179$ whereas for $\lambda\geq2$, for
each $\rho_c$ one has only one such curve.
Also for the WDF there is only one closed curve for each $\rho$ from the maximum
value of $\rho$ up to the curve $\rho=0$.

In Fig.~10 we compare the WDF curves of constant $\rho$ with the SDF curves
for which $\rho=\rho_c$ in the case $\lambda=1$. One notices that the discrepancies between
both curves are very large. In Figs.~11,12 and 13 we make the same comparison
for $\lambda=2,4$ and 10 respectively. 

One finds that for $\rho<0.05$ both curves are quite similar but displaced from
each other. This displacement becomes less pronounced as $\rho$ gets closer to
0.05 so that the best agreement between $\rho$ and $\rho_c$ is reached for
$\rho\approx0.05$. Also as $\lambda$ increases the displacement between both
curves decreases, as it can be seen comparing the oscillators $\lambda=4$ and
$\lambda=10$. For $\rho\stackrel{>}{\sim}0.05$ the curves of fixed $\rho_c$
compared with the curves of fixed $\rho$ contain a certain amount of distortion
which becomes less pronounced as $\lambda$ increases.

Finally we observe that curves with $\rho_c=0.3$ are absent, the largest values
of $\rho_c$ being 0.227, 0.271 and 0.299 respectively for $\lambda=2,4$ and 10,
values which are reached at the origin of phase-space. 

\appendix

\section{Eigenfunctions and eigenvalues of the Morse potential}
\label{sec:a1}

In this Appendix we solve briefly the eigenvalue equation for the Morse
potential (Eq.~(\ref{se2}) of Sec.~\ref{sec:wdfmp})
\begin{equation}
\left[\frac{d^2}{dz^2}-\lambda^2(1-e^{-z})^2+\lambda^2\epsilon_n \right]
\psi_n(z)=0.
\label{a1}
\end{equation}
Making the substitution $y=2\lambda e^{-z}$, Eq.~(\ref{a1}) is written
\begin{equation}
\left[y^2\frac{d^2}{dy^2}+y\frac{d}{dy}
-\lambda^2(1-\frac{y}{2\lambda})^2+\lambda^2\epsilon_n \right]
\psi_n(y)=0,
\label{a2}
\end{equation} 
or,
\begin{equation}
\left[\frac{d^2}{dy^2}+\frac{1}{y}\frac{d}{dy}
+\frac{(\epsilon_n-1)\lambda^2}{y^2}+\frac{\lambda}{y}
-\frac{1}{4} \right]\psi_n(y)=0.
\label{a3}
\end{equation} 

Assuming a solution of the form
\begin{equation}
\psi_n(y)=e^{-\frac{y}{2}}y^su_n(y),
\label{a4}
\end{equation}
and writing $s^2=(\epsilon_n-1)\lambda^2$, from Eq.~(\ref{a3}) we get
\begin{equation}
yu^{\prime\prime}_n+(2s+1-y)u^\prime_n+(\lambda-s-\frac{1}{2})u_n=0,
\label{a5}
\end{equation}
where $u^\prime_n=du_n/dy$ and $u^{\prime\prime}_n=d^2u_n/dy^2$. 
Making $u_n=\sum_{m=0}^na_my^m$, Eq.~(\ref{a5}) may be written as
\begin{equation}
\sum_{m=0}^n[(m+1)(m+2s+1)a_{m+1}+
\left(\lambda-s-m-\frac{1}{2}\right)a_m]y^m=0,
\label{a6}
\end{equation}
which is the differential equation satisfied by the Laguerre generalized
functions.
Eq.~(\ref{a6}) is satisfied if
\begin{equation}
a_{m+1}=\frac{s+m-\lambda+\frac{1}{2} }{ (m+1)(m+2s+1)}\;\;
a_m,
\label{a7}
\end{equation}
which for $m=0$ is
\begin{equation}
a_{1}=\frac{s-\lambda+\frac{1}{2} }{2s+1}\;\;
a_0.
\label{a8}
\end{equation}
Since for $m=n$, $a_{m+1}=0$ and $a_m\not=0$, Eq.~(\ref{a7}) gives
\begin{equation}
s=\lambda-n-\frac{1}{2}.
\label{a9}
\end{equation}
Taking into account Eq.~(\ref{a9}) we get
\begin{equation}
-\epsilon_n=-\frac{2}{\lambda}\left(n+\frac{1}{2}\right)+
\frac{1}{\lambda^2}\left(n+\frac{1}{2}\right)^2,
\label{a10}
\end{equation}
and the recurrence relation for the coefficients of the power series in
Eq.~(\ref{a7}) becomes
\begin{equation}
a_{m+1}=\frac{m-n}{(m+1)(2\lambda-2n+m)}\;\;a_m.
\label{a11}
\end{equation}
Choosing
\begin{equation}
a_0=\frac{(2\lambda-n-1)!}{n!(2\lambda-2n-1)!}\,N(\lambda,n),
\label{a12}
\end{equation}
we obtain
\begin{equation}
a_m=\frac{(-1)^m(2\lambda-n-1)!N(\lambda,n)}{m!(n-m)!(2\lambda-2n-m-1)!},
\label{a13}
\end{equation}
and we get for $\psi_n$ the expression given by Eq.~(33) in
Sec.~\ref{sec:wdfmp}.

\section{Numerical Evaluation of the Wigner function}
\label{sec:a2}

Here we discuss the numerical calculation of the quantity $K_\nu$ defined in
Eq.~(\ref{k}) of Sec.~\ref{sec:wdfmp},
\begin{equation}
K_\nu(\xi)=\frac{1}{2}\int^\infty_0\,
e^{-\frac{\xi}{2}\left(\tau+\frac{1}{\tau}\right)}\;
\tau^\nu\;\frac{d\tau}{\tau}.
\label{kd}
\end{equation}
Here $\nu$ is complex,
\begin{equation}
\nu=N+2ik,
\label{nun}
\end{equation}
where $N$ is an integer and $k$ is the dimensionless momentum $p/\hbar$ and
according to Eqs.(\ref{xi}) and (\ref{ql})
\begin{equation}
\xi=2\lambda e^{-q},\quad q=ax.
\label{xipl}
\end{equation}

Making the transformation
\begin{equation}
\tau=e^u,
\label{tau}
\end{equation}
and considering that only the real part of $K_\nu$ enters into the expression
(51) for the Wigner function we get from Eq.~(\ref{kd}), substituting also $\nu$
according to Eq.~(\ref{nun}), 
\begin{equation}
Re\,[K_\nu(\xi)]=\frac{1}{2}\int^\infty_{-\infty}e^{-\xi \cosh
u}e^{uN}\cos(2ku)du.
\label{yuca4}
\end{equation}

Let us take initially $N=0$, which is the only value needed for the ground state
of the Morse oscillator. Consider also $k\not=0$ as the case $k=0$ is
calculated separately. For convenience we introduce the new variable of
integration
\begin{equation}
z=2ku,
\label{z}
\end{equation}
and use the fact that the integrand is an even function of $u$, obtaining from
Eq.~(\ref{yuca4})
\begin{equation}
Re\,[K_\nu(\xi)]=\frac{1}{2k}\int^\infty_0dz e^{-\xi\cosh(\frac{z}{2k})}\cos z.
\label{yuca5}
\end{equation}

The integral in Eq.~(\ref{yuca5}) is of the form
\begin{equation}
I=\int^\infty_0f(z)\cos zdz,
\label{oba}
\end{equation}
where $f(z)$ is a positive decreasing function  of $z$. In order to avoid
numerical cancellations arising from the change of sign of $\cos(z)$, we replace
this integral by an integral in the interval $[0,\frac{\pi}{2}]$ of a series of
functions. 

We decompose $I$ as follows
\begin{equation}
I=I_1+I_2,
\label{ii}
\end{equation}
where 
\begin{equation}
I_1=\int^{\frac{\pi}{2}}_0f(z)\cos zdz,
\label{i1}
\end{equation}
and
\begin{equation}
I_2=\int^\infty_{\frac{\pi}{2}}f(z)\cos zdz=-\int^\infty_0f(y+\frac{\pi}{2})\sin
ydy,
\label{i2}
\end{equation}
where we made the substitution $z=y+\frac{\pi}{2}$ in $I_2$. The interval of
integration of the integral $I_2$ is  divided in the set of intervals $[2\pi
s,2\pi(s+1)]$, $s=0,1,\cdots$ so that $I_2$ is written
\begin{equation}
I_2=-\sum^\infty_{s=0}\int^{2\pi(s+1)}_{2\pi s}f(y+\frac{\pi}{2})\sin
ydy.
\label{i22}
\end{equation}

Making now the substitution
\begin{equation}
z=y-2\pi s,
\label{s}
\end{equation}
for the integral in the interval $[2\pi s,2\pi(s+1)]$, one obtains for
Eq.~(\ref{i22})
\begin{equation}
I_2=-\sum^\infty_{s=0}\int^{2\pi}_0f(z+(2s+\frac{1}{2})\pi)\sin
zdz.
\label{i23}
\end{equation}

However $\sin z$ changes sign in the interval $[0,2\pi]$ which leads to
cancellation errors if $f$ is slowly varying in the interval. In order to
eliminate the oscillation of sign of the integrand we decompose again the
integral in the intervals $[0,\pi]$ and $[\pi,2\pi]$ obtaining
\begin{equation}
I_2=-\sum^\infty_{s=0}\left\{\int^{\pi}_0\left[f\left(z+\left(2s+\frac{1}{2}
\right)\pi\right) -f\left(z+\left(2s+\frac{3}{2}\right)\pi\right)\right] \sin
zdz\right\}.
\label{i24}
\end{equation}

As $f$ is assumed to be monotonically decreasing each integrand in
Eq.~(\ref{i24}) is now always positive. As the domain of integration of the
integral $I_1$ is the interval $[0,\frac{\pi}{2}]$ we make the translation
$v=z-\frac{\pi}{2}$, obtaining
\begin{equation}
I_2=-\sum^\infty_{s=0}\int^{\frac{\pi}{2}}_{-\frac{\pi}{2}}
[f(v+(2s+1)\pi) -f(v+(2s+2)\pi)] \cos vdv.
\label{i25}
\end{equation}

Making $v\to-v$ for the contribution from the interval $[-\frac{\pi}{2},0]$ in
Eq.~(\ref{i25}) and adding the contribution from $I_1$ one gets finally
\begin{eqnarray}
\int^\infty_0f(x)\cos xdx&=&\int^{\frac{\pi}{2}}_{0}dx\cos x\{
f(x)-\sum^\infty_{s=0}[f((2s+1)\pi-x)+f((2s+1)\pi+x)
 \nonumber \\
&-&f((2s+2)\pi-x) -f((2s+2)\pi+x) ]\}.
\label{i2f}
\end{eqnarray}

For $f(x)$ monotonic each term of the series contributes with the same sign, 
however errors may arise from the
subtraction of the sum of the series from $f(x)$.

In the general case in which $N\geq0$ in Eq.~(\ref{yuca4}), the function $f(z)$
in Eq.~(\ref{oba}) becomes  
\begin{equation}
f_N(z)=\frac{1}{2k}e^{-\xi\cosh\frac{z}{2k}}\cosh\left(\frac{Nz}{2k} \right),
\;\;z\geq0,\;N=0,1,\cdots
\label{b18}
\end{equation}

The function $f_N(z)$ has $M\,(M\!\!\leq \!\!N)$ extremes at the real
positive points $z^{(N)}_1,z^{(N)}_2,\cdots, z^{(N)}_M$, labeled in the order of
increasing magnitude. This function decreases monotonically for $z>z^{(N)}_M$
and it may still be useful to apply the decomposition (\ref{i2f}) in order to
eliminate errors due to cancellations of contributions of opposite sign from the
integrand $f_N(z)\cos z$. In the general case only the sign of the first $s_N$
terms of the series in Eq.~(\ref{i2f}) may oscillate, where $s_N$ is given
roughly by the smallest integer satisfying $(2s_N+1)\pi>z^{(N)}_M$. For $s>s_N$
the sign of the terms of the series in Eq.~(\ref{i2f}) is always positive.

In fact, one may determine the extremes of $f_N(z)$ by expressing $f_N(z)$ in
terms of $\cosh(z/2k)$ and applying the condition $\partial f_N(z)/\partial
z=0$. By making this substitution one gets for Eq.~(\ref{b18}) the expression
\begin{equation}
f_N(z)=\frac{1}{2k}e^{-\xi y}\sum^{[N/2]}_{j=0}A_j(-1)^jy^{N-2j},\;\;
y=\cosh(\frac{z}{2k}),
\label{f}
\end{equation} 
where $[N/2]$ denotes the largest integer contained in $N/2$ and
\begin{equation}
A_j=\sum^{[N/2]}_{i=j}\left(\begin{array}{c}
N\\2i\end{array}
\right)\left(\begin{array}{c}
i\\j\end{array}
\right).
\label{aa}
\end{equation}

Thus we get the extrema of $f_N(z)$ as the roots of a polynomial of 
degree $N$.

For $N=1$ the maximum occurs at 
\begin{equation}
z^{(1)}_1=2k\cosh^{-1}\, \left(\frac{1}{\xi}\right).
\label{ma1} 
\end{equation}

For $N=2$ the maximum will be at

\begin{equation}
z^{(2)}_1=2k\cosh^{-1} \left(\frac{2}{\xi}+\sqrt{\frac{4}{\xi^2}+2}\;\right).
\label{ma2} 
\end{equation}

\vglue 0.01cm
\begin{figure}[ht]
\begin{center}
\vglue 1.5cm
\mbox{\epsfig{file=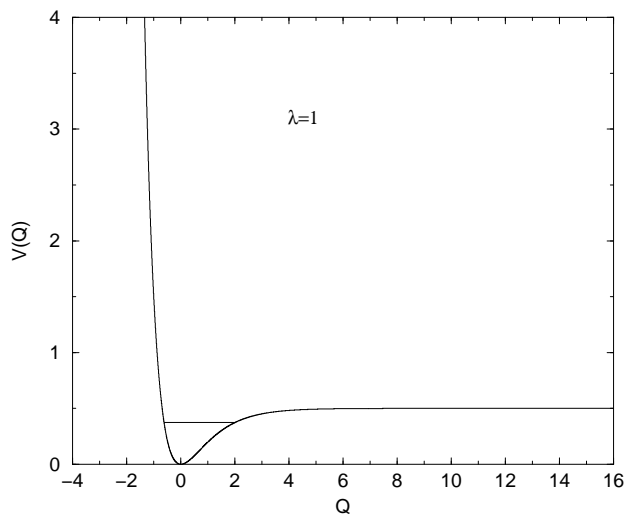,width=0.5\textwidth,angle=0}}   
\end{center}
\caption{ The Morse potential $V(Q)$ defined by Eq.~(39c) 
and the corresponding bound-state energies $E_n$ given by Eq.~(39d), for 
$\lambda=1$.}
\label{fig1}
\end{figure}
\begin{figure}[ht]
\begin{center}
%
\vglue 1.5cm
\mbox{\epsfig{file=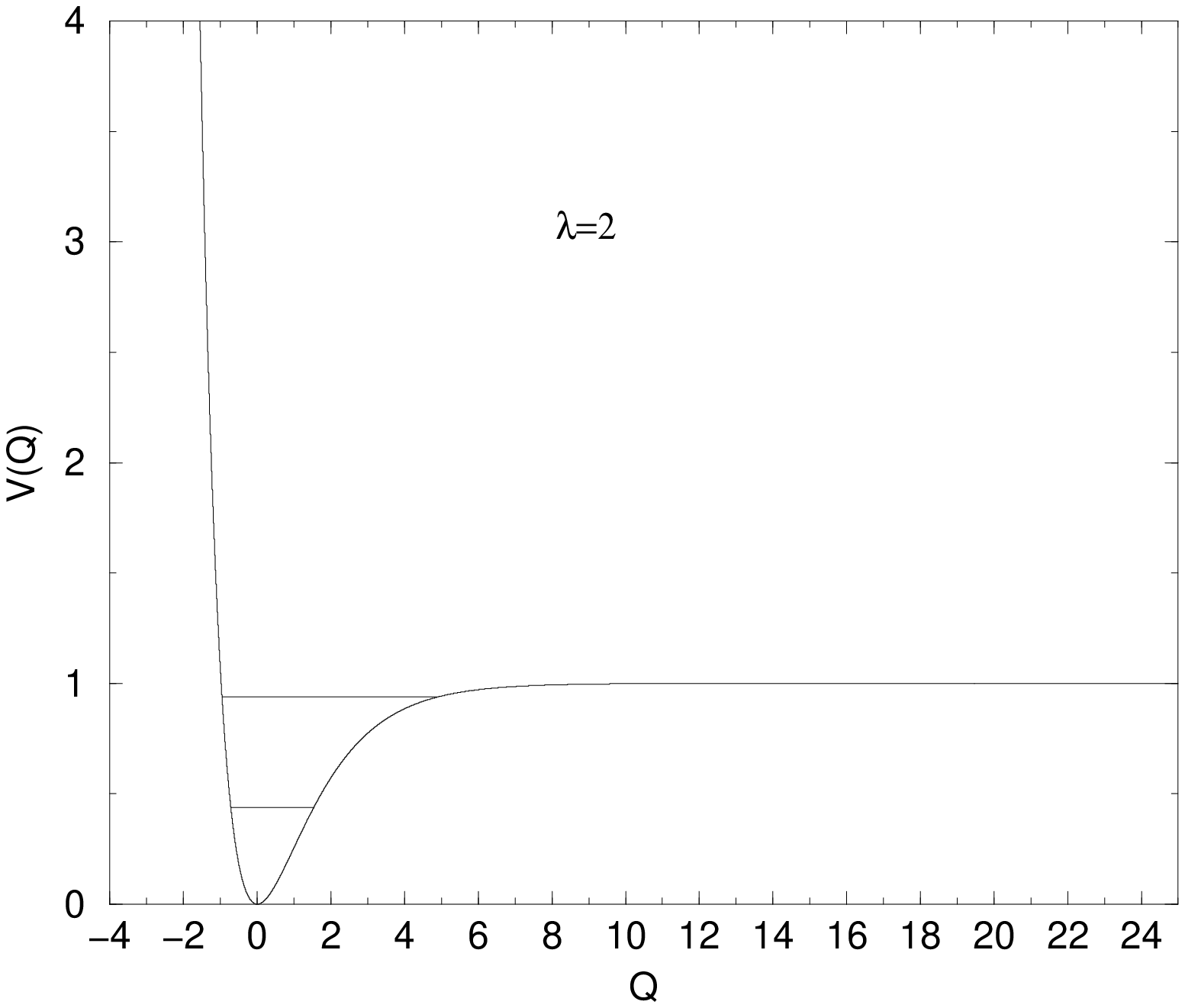,width=0.5\textwidth,angle=0}}   
\end{center}
\caption{Same as Fig.~1 for $\lambda=2$.}
\label{fig2}
\end{figure}

\begin{figure}[ht]
\begin{center}
%
\vglue 1.5cm
\mbox{\epsfig{file=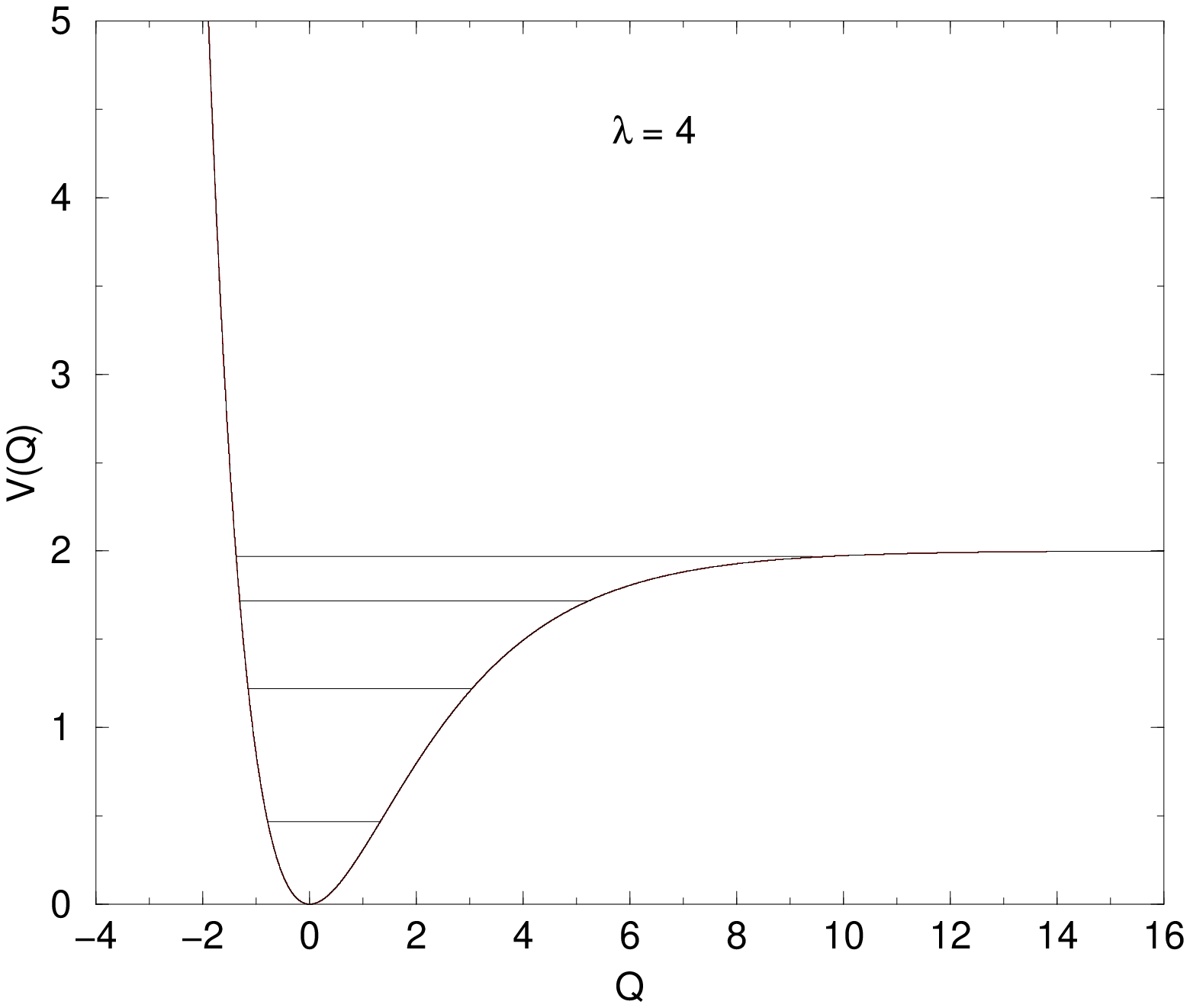,width=0.5\textwidth,angle=0}}   
\end{center}
\caption{Same as Fig.~1 for $\lambda=4$.}
\label{fig3}
\end{figure}

\begin{figure}[ht]
\begin{center}
%
\vglue 1.5cm
\mbox{\epsfig{file=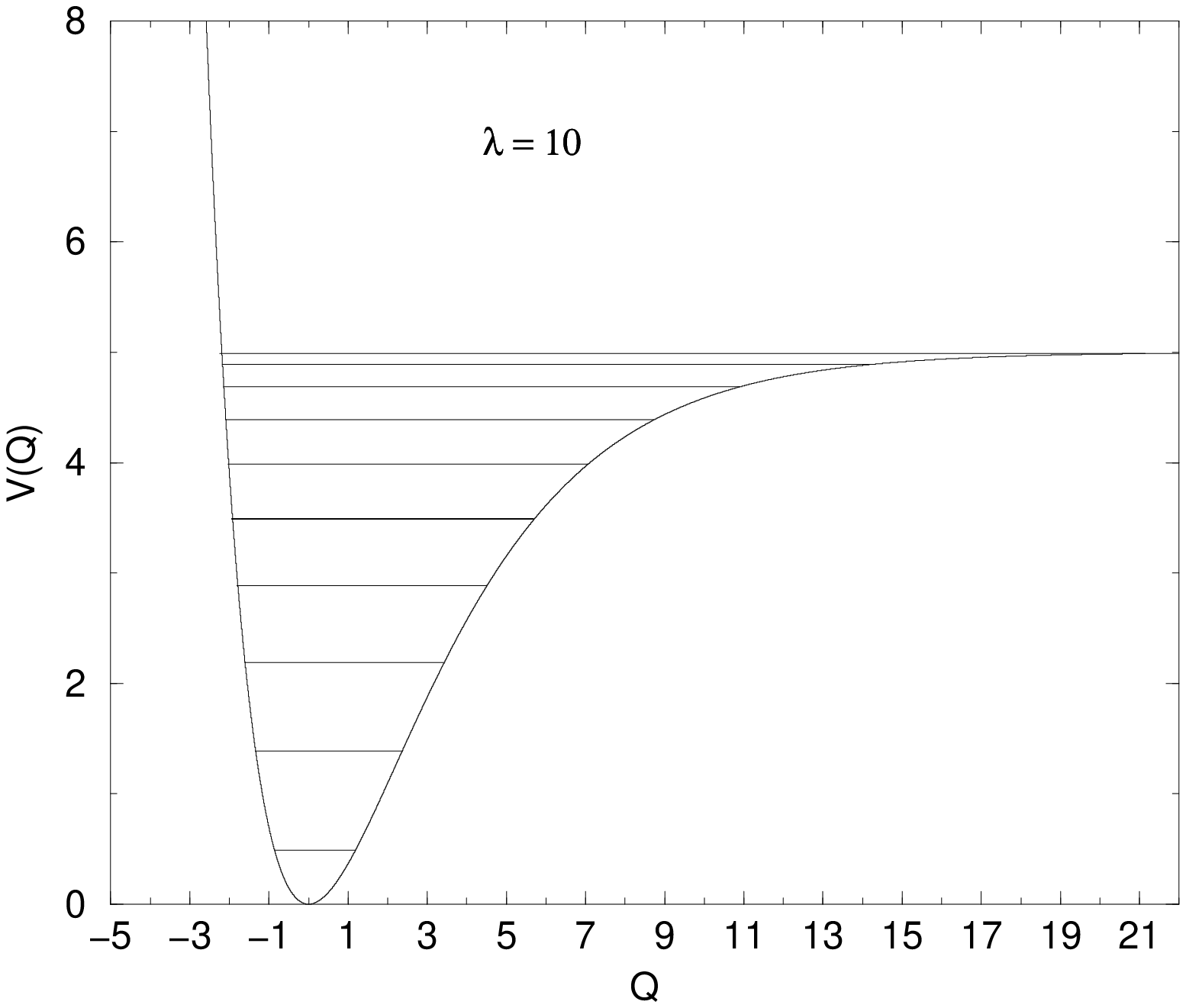,width=0.5\textwidth,angle=0}}   
\end{center}
\caption{Same as Fig.~1 for $\lambda=10$.}
\label{fig4}
\end{figure}

\begin{figure}[ht]
\begin{center}
%
\vglue 1.5cm
\mbox{\epsfig{file=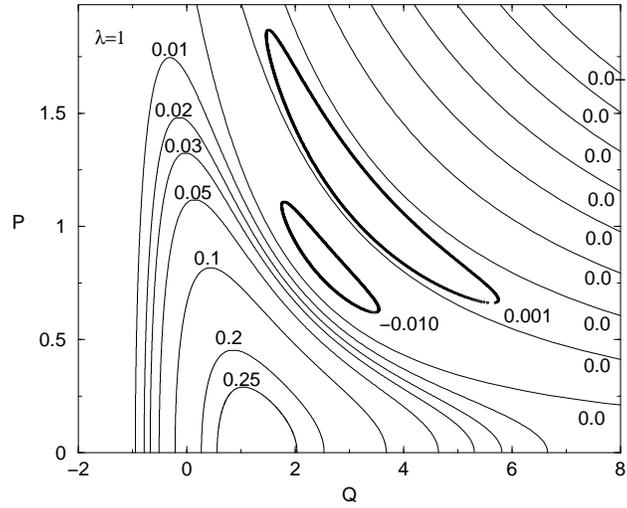,width=0.5\textwidth,angle=0}}   
\end{center}
\caption{Curves of constant Wigner distribution function $\rho(Q,P)$
in units of $\hbar^{-1}$ for the ground state of the Morse oscillator with
$\lambda=1$.}
\label{fig5}
\end{figure}

\begin{figure}[ht]
\begin{center}
%
\vglue 1.5cm
\mbox{\epsfig{file=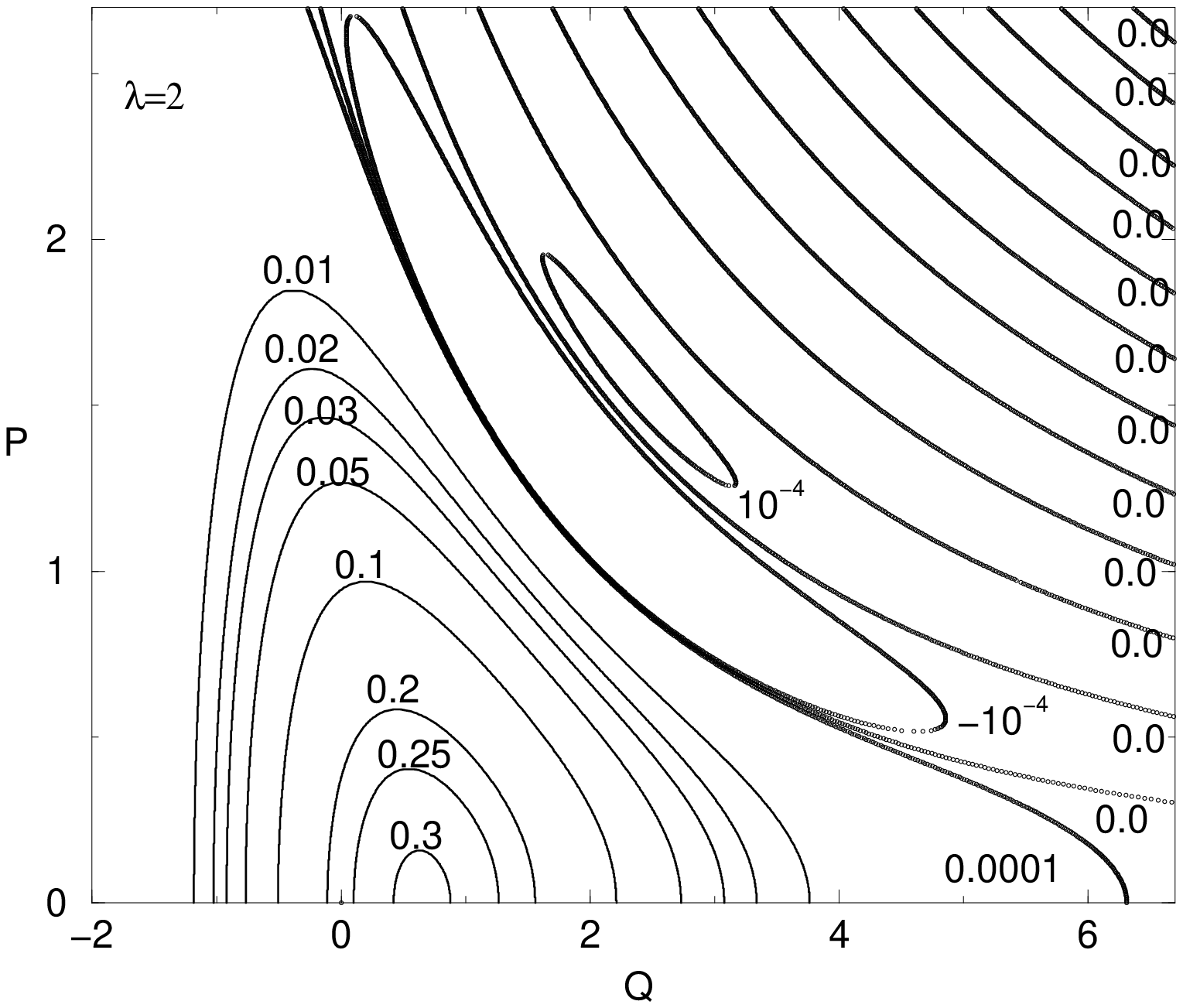,width=0.5\textwidth,angle=0}}   
\end{center}
\caption{Same as Fig.~5 for $\lambda=2$.}
\label{fig6}
\end{figure}

\begin{figure}[ht]
\begin{center}
%
\vglue 1.5cm
\mbox{\epsfig{file=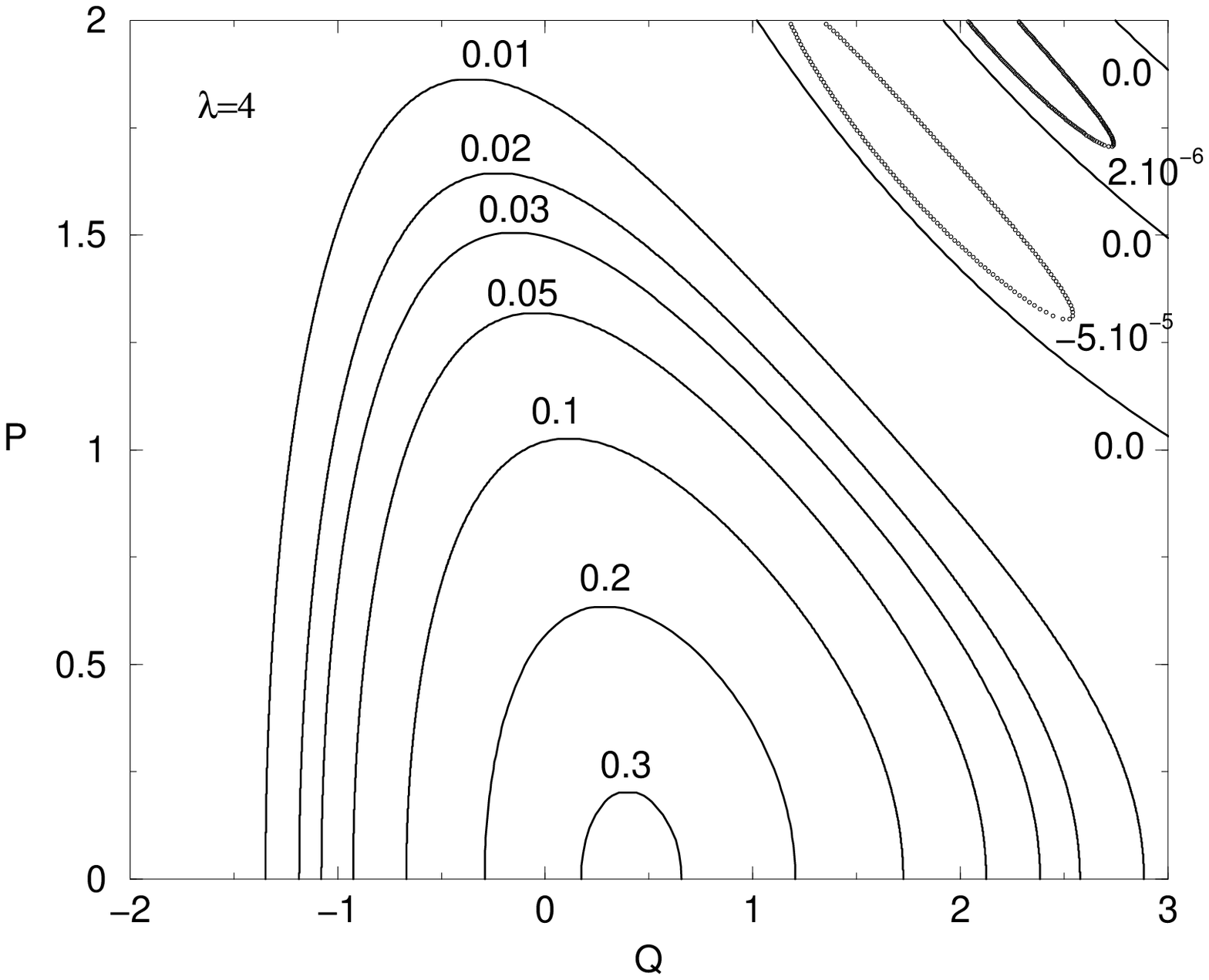,width=0.5\textwidth,angle=0}}   
\end{center}
\caption{Same as Fig.~5 for $\lambda=4$.}
\label{fig7}
\end{figure}

\begin{figure}[ht]
\begin{center}
%
\vglue 1.5cm
\mbox{\epsfig{file=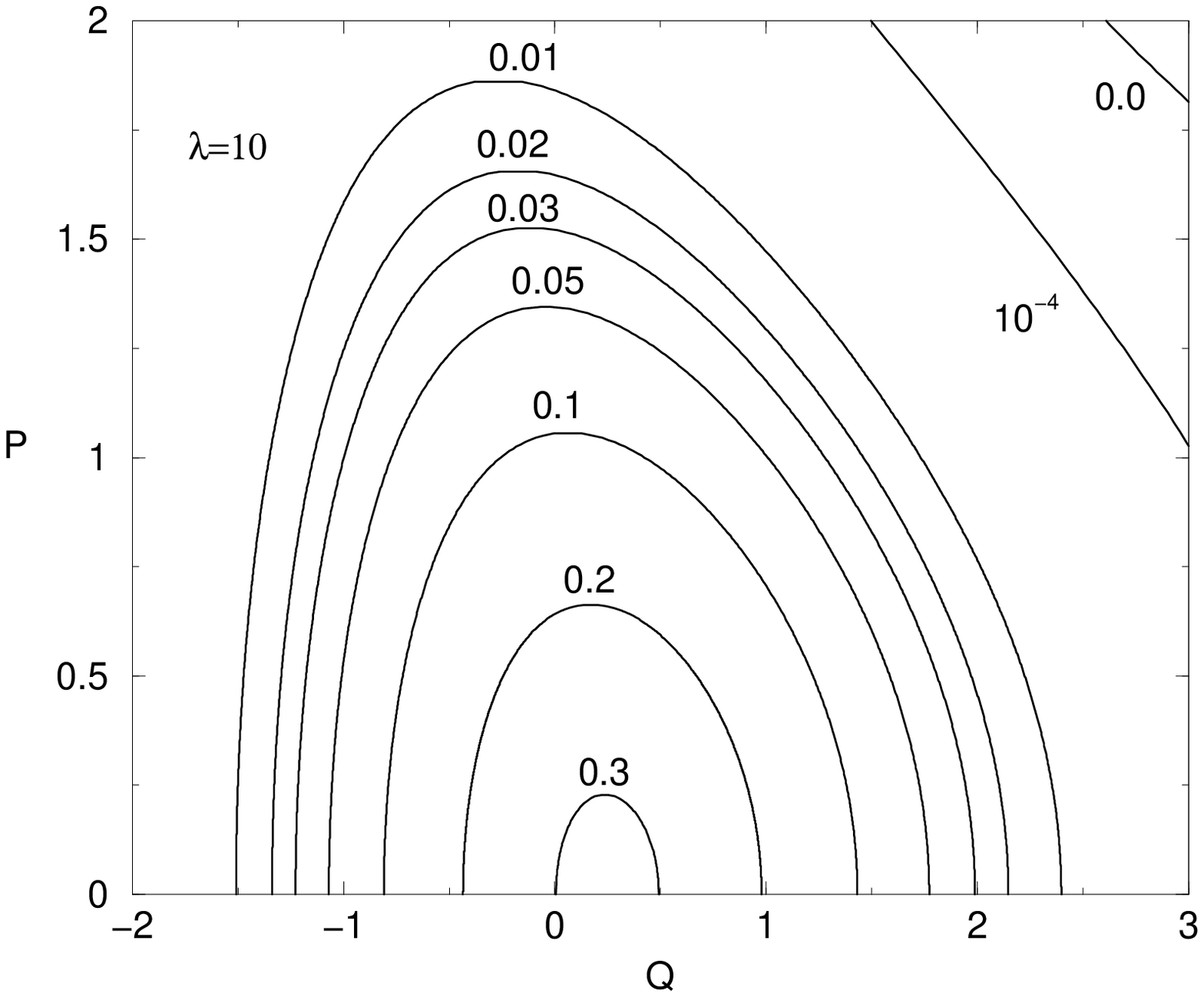,width=0.5\textwidth,angle=0}}   
\end{center}
\caption{Same as Fig.~5 for $\lambda=10$.}
\label{fig8}
\end{figure}

\begin{figure}[ht]
\begin{center}
%
\vglue 1.5cm
\mbox{\epsfig{file=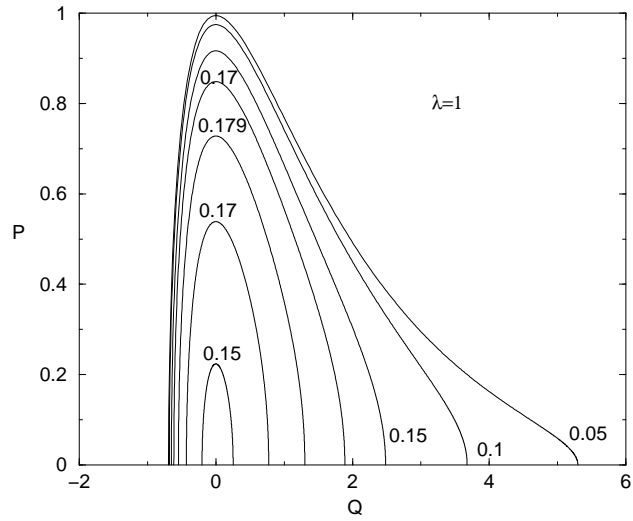,width=0.5\textwidth,angle=0}}   
\end{center}
\caption{Curves of constant semi-classical
distribution function $\rho_c(Q,P)$ in units of $\hbar^{-1}$ for the ground
state of the Morse oscillator with $\lambda=1$. For $0.145<\rho_c<0.179$ one
has two curves for each value of $\rho_c$.}
\label{fig9}
\end{figure}

\newpage
\begin{figure}[ht]
\begin{center}
%
\vglue 1.5cm
\mbox{\epsfig{file=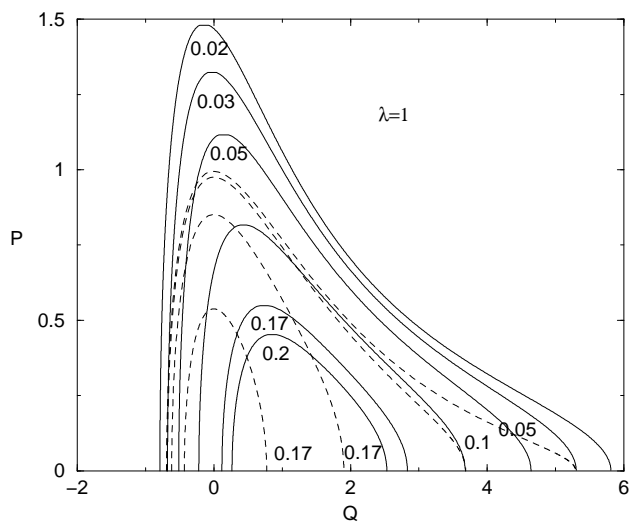,width=0.5\textwidth,angle=0}}   
\end{center}
\caption{Full curves are curves with a constant value of the Wigner
distribution function $\rho$ and dashed lines have constant value of the
semiclassical distribution function $\rho_c$ for the ground state of the Morse
oscillator with $\lambda=1$. Except for $\rho=0.2$, for each full line with a
given $\rho$ one has one or two  
corresponding dashed lines with $\rho_c$ such that $\rho_c=\rho$.}
\label{fig10}
\end{figure}

\begin{figure}[ht]
\begin{center}
%
\vglue 1.5cm
\mbox{\epsfig{file=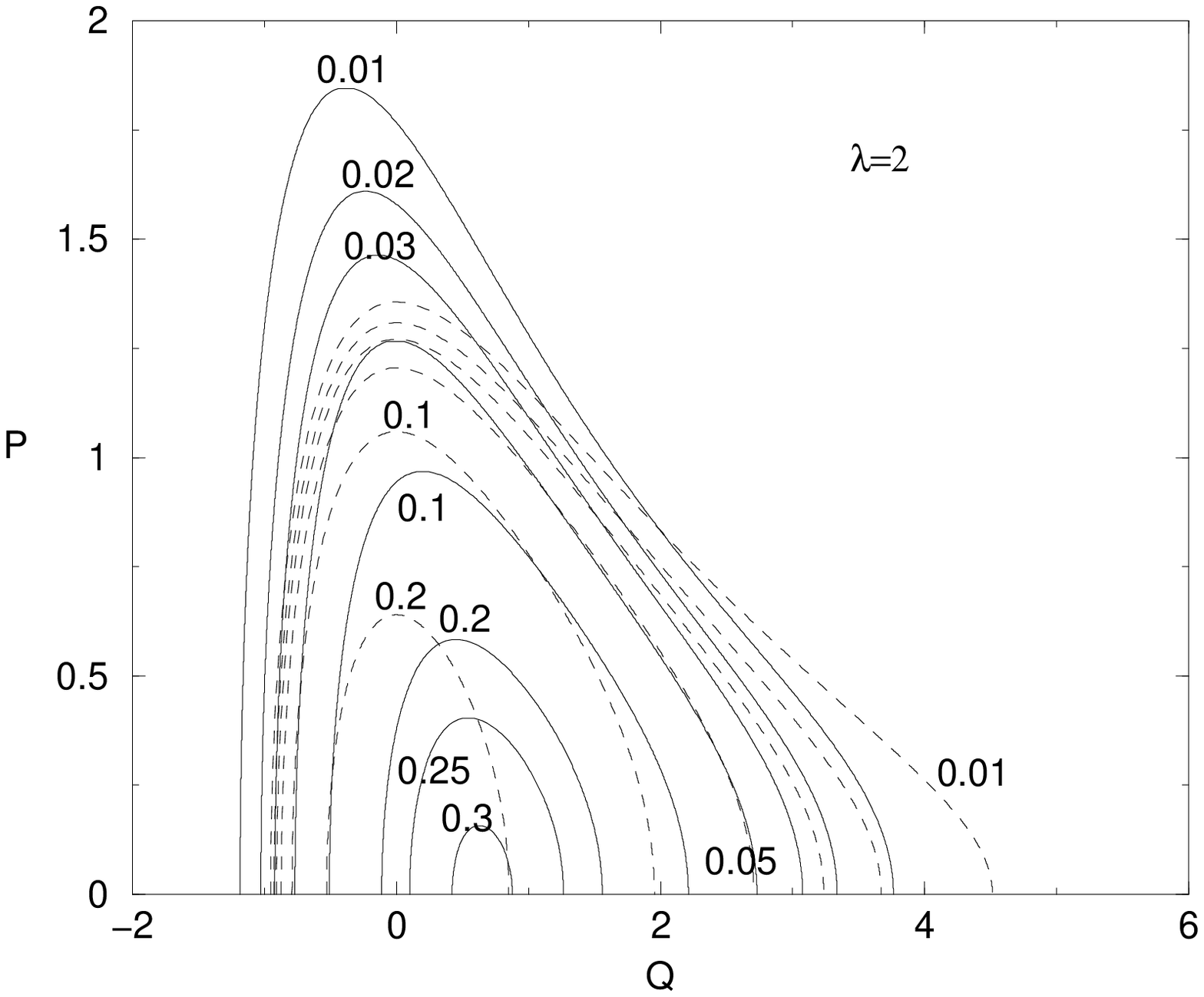,width=0.5\textwidth,angle=0}}   
\end{center}
\caption{Same as Fig.~10 for $\lambda=2$. Except for $\rho=0.3$ and $\rho=0.25$,
for each full line one has a corresponding dashed line with $\rho_c$ such that
$\rho_c=\rho$.}
\label{fig11}
\end{figure}

\begin{figure}[ht]
\begin{center}
%
\vglue 1.5cm
\mbox{\epsfig{file=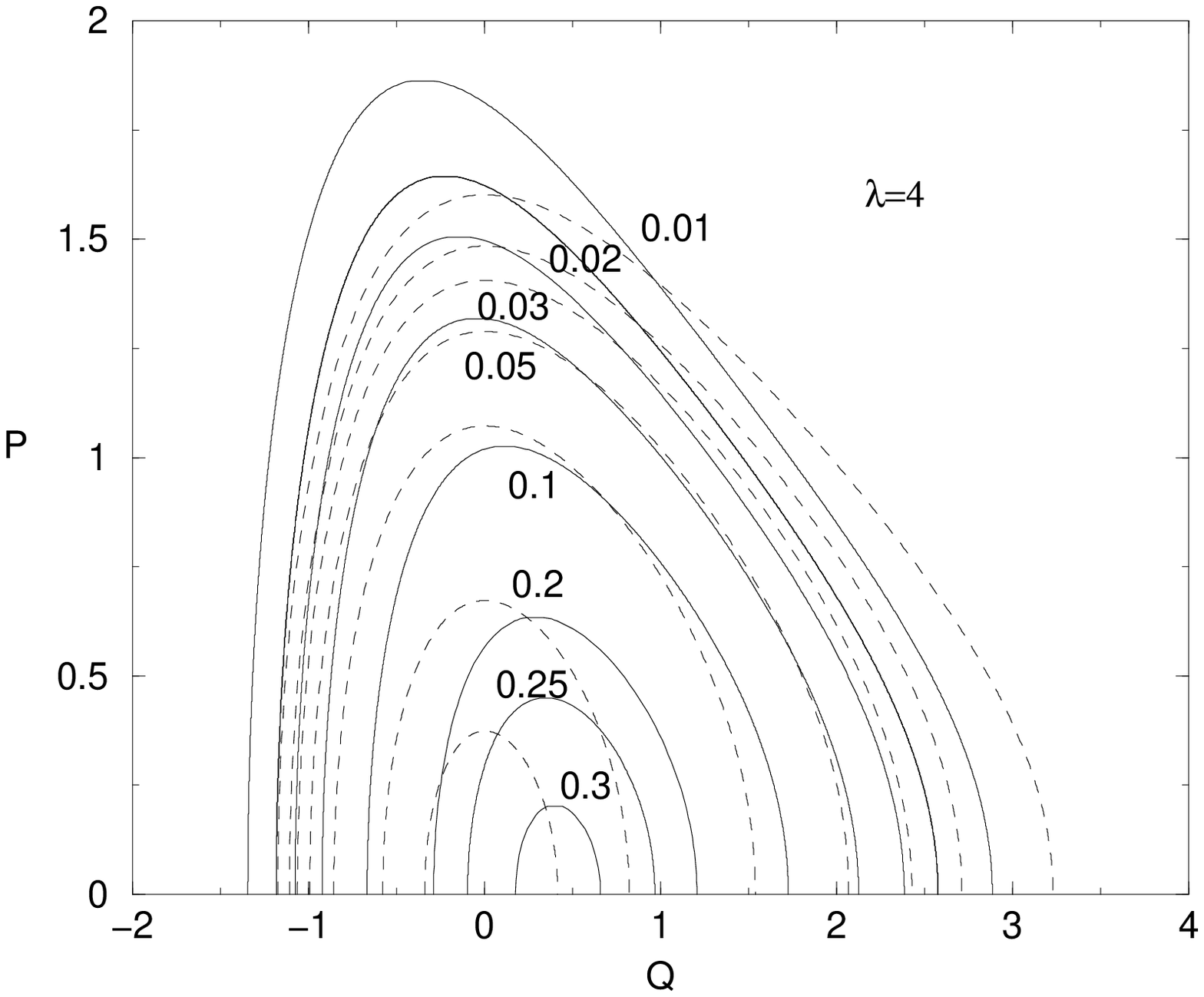,width=0.5\textwidth,angle=0}}   
\end{center}
\caption{Same as Fig.~10 for $\lambda=4$. Except for $\rho=0.3$, for each full
line one has a corresponding dashed line with $\rho_c$ such that $\rho_c=\rho$.}
\label{fig12}
\end{figure}

\begin{figure}[ht]
\begin{center}
%
\vglue 1.5cm
\mbox{\epsfig{file=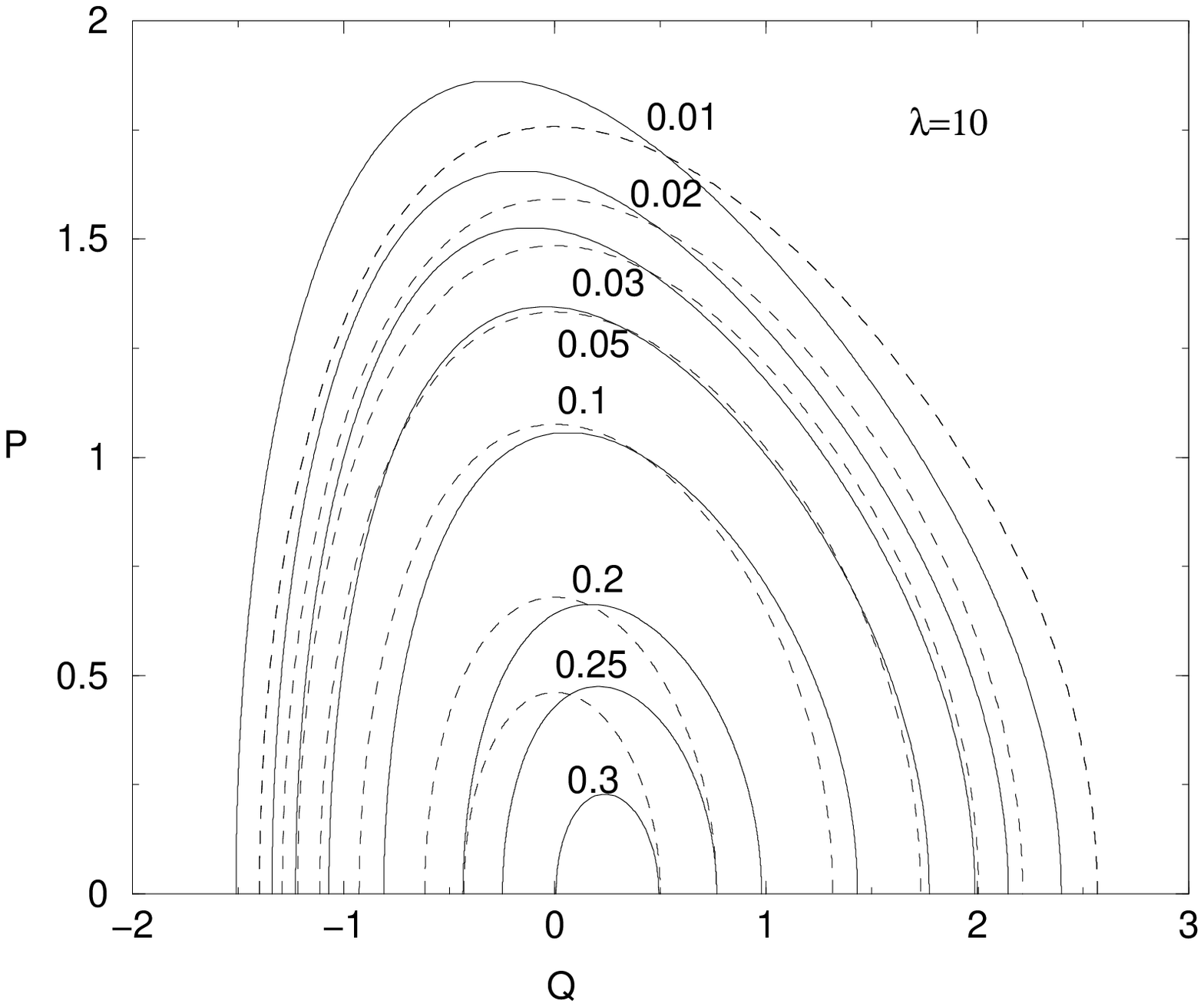,width=0.5\textwidth,angle=0}}   
\end{center}
\caption{Same as Fig.~10 for $\lambda=10$. Except for $\rho=0.3$, for each full
line one has a corresponding dashed line with $\rho_c$ such that $\rho_c=\rho$.}
\label{fig13}
\end{figure}


\end{document}